\documentclass[12pt]{iopart}

\usepackage{iopams}
\usepackage{graphicx}

\newcommand{\beq}{\begin{eqnarray}}
\newcommand{\eeq}{\end{eqnarray}}

\def\bxi{\mbox{\boldmath$\xi$}}

\def\btau{\mbox{\boldmath$\tau$}}
\def\br{\mbox{\boldmath$r$}}
\def\bn{\mbox{\boldmath$n$}}
\def\bk{\mbox{\boldmath$k$}}
\newcommand{\bm}[1]{\mbox{\boldmath$#1$}}
\begin{document}

\title{Effects of crosslinks on motor-mediated filament organization}

\author{Falko Ziebert$^{1}$, Igor S. Aranson$^{1}$,  and Lev S. Tsimring$^2$}
\ead{aronson@msd.anl.gov}
\address{$^1$ Materials Science Division, Argonne National Laboratory,
9700 S Cass Avenue, Argonne, Illinois 60439, USA}%
\address{$^2$ Institute for Nonlinear Science, University of California, San Diego, La Jolla, California 92093-0402, USA}%

\begin{abstract}
Crosslinks and molecular motors play an important role in the organization of cytoskeletal
filament networks. Here we incorporate the effect of crosslinks into our
model of polar motor-filament organization [Phys. Rev. E {\bf 71}, 050901 (2005)],
through suppressing the relative sliding of filaments in the course of motor-mediated alignment.
We show that this modification leads to a nontrivial macroscopic behavior, namely
the oriented state exhibits a transverse
instability in contrast to the isotropic instability that occurs without crosslinks.
This transverse instability leads to the formation of dense extended bundles of oriented filaments,
similar to recently observed structures in actomyosin.
This model also can be applied to situations with two oppositely directed
motor species or motors with different processing speeds.
\end{abstract}

%Uncomment for PACS numbers title message
%\pacs{00.00, 20.00, 42.10}
% Keywords required only for MST, PB, PMB, PM, JOA, JOB?
%\vspace{2pc}
%\noindent{\it Keywords}: Article preparation, IOP journals
% Uncomment for Submitted to journal title message
%\submitto{\JPA}
% Comment out if separate title page not required
\maketitle

%subsections are possible
%%%%%%%%%%%%%%%%%%%%%%%%%%%%%%%%%%%%%%%%%%%%%%%%%%%%%%%%
\section{\label{Intro}Introduction}
%%%%%%%%%%%%%%%%%%%%%%%%%%%%%%%%%%%%%%%%%%%%%%%%%%%%%%%%

Biological cells consist to a large degree of
a complex, self-organizing viscoelastic fluid, the cytosol.
Its main constituents
include cytoskeletal proteins such as actin and tubulin, which exist mainly in the
polymerized form as semi-flexible actin filaments and stiff microtubules
\cite{Lodish:1999,Alberts:2001}.
The entangled networks of microtubules and actin filaments
form the cytoskeleton of most eukaryotic cells, stabilize their morphology
and  determine the rheological properties of the cytosol.
These intricate networks are created and maintained
by an efficient mechanism which involves
various types of motor proteins as well as passive crosslinking proteins.
Motors are specialized protein molecules that
move along the cytoskeletal polymer scaffold
and perform directed intracellular transport \cite{Howard:2001}.
Additionally, if motors bind to more than one filament, they are able to move
the filaments and reorganize the cytoskeleton itself.
The crosslinks connect different filaments but do not move along them. Their main
function is thus believed to provide rigidity and elasticity to the cytoskeleton.

Various experiments have been performed
in recent years that shed light on the viscoelastic behavior
of entangled cytoskeletal filament solutions, ranging from
filament-motor mixtures \cite{LeGoff:2002.1}, crosslinked filaments
\cite{Janmey:1995,Janmey:2005,Weitz:2006.1}, and systems of filaments,
motors and crosslinks \cite{MacKintosh:2007.1}.
Surprising  new effects have been found such as active fluidization of actin gels
by myosin motors \cite{Kaes:2002.1}.

Maintained in a state far from equilibrium,
the active filaments exhibit a strong tendency towards self-organization.
Bundles and contracting states have been found {\em in vitro} in actomyosin extracted from muscle cells
\cite{Takiguchi:91.1}, and various patterns like ray-like asters, spindle-like structures and rotating vortices
have been reported in quasi two-dimensional mixtures of microtubules and motors
\cite{Nedelec:1997.1,Leibler:2001.1}.
These dissipative structures have inspired many theoretical efforts
\cite{Sekimoto:1996.1,Kardar:2001.1,Liverpool:2003.2,Kruse:2004.2,Tsimring_Aranson,Ziebert:2005.1,Tsimring_Aranson2}
directed towards modeling active filament solutions.

While crosslinks so far have been mainly investigated only in the context of rheology,
recently their influence on the dynamics and
self-organization also attracted attention \cite{Smith:06.1}.
In particular, it was shown that crosslinks facilitate the formation of bundles in the actin-myosin system:
at high concentration of adenosine triphosphate (ATP),  actin-myosin systems display an isotropic phase;
in the course of depletion of ATP however, myosin motors become static
crosslinks and initiate the formation
of oriented bundles and cluster-like patterns. Reintroduction of ATP in the
bundled state resulted in consequent dissolution of the structures and reestablishment of the isotropic state.

Motivated by these experimental results, we focus here on the
effects of static crosslinks on the self-organization of polar filaments and
generalize the model for microtubule-motor interaction introduced
in Refs.~\cite{Tsimring_Aranson,Tsimring_Aranson2}.
In that model, the complicated process of filament interaction via multi-headed molecular motors
was approximated by instant binary ``inelastic collisions'', leading to alignment of the filament
orientation vectors and attraction between their centers of mass. Crosslinks alter these
interaction rules. In particular, if two parallel filaments are
cross-linked, they are not able to slide past each other and
become collocated. We model this effect here by suppressing  relative sliding of the filaments in the
course of alignment. Our analysis shows that this relatively minor modification produces
a nontrivial macroscopic effect, namely the isotropic density instability of the polar oriented state of the filaments becomes transverse.
In the nonlinear regime, this new kind of instability leads to the formation of dense oriented bundles,
similar to those seen in experiments \cite{Smith:06.1}. In contrast, the model without crosslinks
demonstrates an isotropic instability in which  density and orientation
of the filaments are uncorrelated, and no bundling occurs.

%%%%%%%%%%%%%%%%%%%%%%%%%%%%%%%%%%%%%%%%%%%%%%%%%%%%%%%%
\section{\label{model}Model}
%%%%%%%%%%%%%%%%%%%%%%%%%%%%%%%%%%%%%%%%%%%%%%%%%%%%%%%%
Here we outline the model of self-organization of microtubule-motor mixtures developed in our earlier works,
Refs.~\cite{Tsimring_Aranson,Tsimring_Aranson2}.
The microtubules are modeled as identical rigid polar rods of length $L$, and the molecular motors are introduced
implicitly through corresponding interaction probabilities.
Binary interactions of microtubules via multi-headed molecular motors are approximated by instant inelastic collisions
leading to alignment of the microtubule orientation angles $\phi_{1,2}$ (or, equivalently, the unit vectors
$\bn_{1,2}=(\cos \phi_{1,2}, \sin \phi_{1,2})$) according to the following rules:
 \begin{equation}
\left(\begin{array}{c} \phi^a_1 \\ \phi^a_2
\end{array}\right) = \left( \begin{array}{cc} \gamma   & 1-\gamma  \\
1-\gamma &  \gamma \end{array} \right) \left(\begin{array}{c} \phi_1
\\ \phi_2
\end{array}\right)\,.
\label{collis}
\end{equation}
Here $\phi_{1,2}^a$ are the orientations of the two rods after the collision, and the constant ``restitution'' parameter $\gamma$
characterizes the inelasticity of the collision (in analogy to the restitution
coefficient in granular media). The angle between the two rods is
reduced after the collision by the ``inelasticity''  factor
$\varepsilon=2\gamma-1$. Of special interest is the totally inelastic collision corresponding  to $\gamma=1/2$ or
$\varepsilon=0$. In this case the
rods acquire the same orientation along the bisector
$\bar {\bn} = (\cos \bar \phi, \sin \bar \phi)$, and their center of mass positions, $\br_{1,2}$, also align:
 \begin{eqnarray}
\label{wo_crossl_phi}
\phi^a_1& =& \phi_2^a  =  \bar \phi=\frac{\phi_1+\phi_2}{2}\,,\\
{\br }_{1}^a & =& \br _{2}^a  = \bar {\br} =\frac{ \br_1 + \br_2}{2}\,.
\label{wo_crossl}
\end{eqnarray}
Here  $\phi^a_{1,2}$  and $\br^a_{1,2}$  are the orientation angles and the center of mass positions
after the collision. We assume that the alignment through inelastic interaction occurs only if the initial angle difference
$|\phi_1-\phi_2|$ is smaller than a certain
maximum interaction angle $\phi_0$. For $|\phi_1-\phi_2|>\phi_0$, the angles and the positions are unchanged.
The analysis of Refs. \cite{Tsimring_Aranson,Tsimring_Aranson2} showed that in the spatially homogeneous case,
the rods exhibited a spontaneous orientation transition if the density of the motors (or of the filaments)
exceeded a critical density. Furthermore, for even higher densities, another instability was predicted which
is isotropic and leads to inhomogeneous density variations.

The dynamics of this model can be described by the master equation for the
probability distribution function $P(\br,\phi,t)$
to find a rod at position $\br$ with orientation $\bn=(\cos \phi, \sin \phi)$:
\begin{eqnarray}
&&\frac{\partial P({\br},\phi,t)}{\partial t}=
\frac{\partial^2P(\br,\phi,t)}{\partial \phi^2}+ \partial_i D_{ij}
\partial_j P(\br, \phi,t)+\mathcal{I}(\br,\phi,t)\,.
\label{master}
\end{eqnarray}
The first two terms on the right hand side describe rotational and translational diffusion,
with an anisotropic diffusion matrix of the form
\begin{eqnarray}
D_{ij}=\frac{1}{D_r}[D_\parallel n_i n_j + D_\perp (\delta_{ij}-n_i
n_j)]\,.
\label{diff}
\end{eqnarray}
The rotational, $D_r$, parallel, $D_\parallel $, and perpendicular, $D_\perp$, diffusion coefficients
for rigid rods in a viscous fluid are well known \cite{DoiEd}.
The third term in Eq.~(\ref{master}) is the collision integral,
\begin{eqnarray}
\hspace{-1cm}\mathcal{I}(\br,\phi,t)&=&
\int\int d\br_1 d\br_2 \int_{-\phi_0}^{\phi_0} d \phi_1 d \phi_2
 W(\br_1-\br_2,\bn_1, \bn_2)
P(\br_1,\phi_1)P(\br_2,\phi_2) \nonumber \\
&\times& \left[ \delta (\phi -\phi_1^a)
\delta\left(  {\br } - {\br}^a_1 \right)-
\delta (\phi - \phi_1)
\delta\left(\br-\br_1\right)\right]\,, \label{coll_int}
\end{eqnarray}
where the localization of spatial interactions %defined by Eqs.~(\ref{wo_crossl})
is introduced through a certain probabilistic kernel $W(\br_1-\br_2, \bn_1, \bn_2)$
\cite{Tsimring_Aranson,Tsimring_Aranson2}.

The kernel $W$, expressing the probability of interaction between the rods as a function of the distance between their midpoints and their orientations,
can be obtained from the following conditions: (i) since the size of motors is small compared to the length of filaments,
two rods interact only if they intersect; (ii) due to translational and rotational invariance, the kernel depends only on differences $\phi_1-\phi_2$ and
$\br_1 - \br_2$; (iii) the kernel is invariant with respect to permutations $\bn_1 \to \bn_2$, $\br_1 \to \br_2$.
The kernel can be represented as a product of two parts: a part $W_0$ which accounts for spatial localization due to the overlap condition
of the filaments, and a part describing the motor-induced collision anisotropy.

The first part can be derived from the intersection condition between two rods  with orientations ${\bn}_{1,2}$.
It is easy to verify that the rods overlap if
\begin{eqnarray}
 | (\br_1-\br_2) \times \bn_1 | &\le& L |\bn_1 \times \bn_2 | /2\,, \\
| (\br_1-\br_2) \times \bn_2 | &\le& L |\bn_1 \times \bn_2 | /2\,
\end{eqnarray}
holds.
This overlap condition can be expressed in terms of discontinuous $\Theta$-functions,
\begin{equation}
\hspace{-1.8cm} W_0  = W_n \Theta( L |\bn_1 \times \bn_2 | -2| (\br_1-\br_2) \times \bn_1 |)
\Theta( L |\bn_1 \times \bn_2 | -2| (\br_1-\br_2) \times \bn_2 | )\,,
\label{rhomb}
\end{equation}
where $W_n$ is a normalization 
constant, so that $\int W_0 d\br=1$.
Since this discontinuous kernel is difficult  for calculations, the  $\Theta$-functions can be approximated by smooth
Gaussians %$\exp(-x^2) $
yielding
\begin{equation}
\hspace{-1.4cm}W_0(\br_1-\br_2,\bn_1,\bn_2 ) \sim %{1\over b^2\pi \sin^2 (\phi_1-\phi_2)  }
\exp\left[-4{\left(({\br}_1-\br_2)  \times {\bn}_1\right)^2+ \left(({\br}_1-\br_2)
\times {\bn}_2\right)^2\over b^2 |\bn_1 \times \bn_2 |^2  }\right] \label{kernelb2}\,,\,\,
\end{equation}
where $b$ is a cutoff length of order $b \lesssim L$.
It is convenient to transform the kernel
to the following representation (the integral of the kernel is normalized to $1$):
\begin{equation}
\hspace{-1.6cm}  W_0({\bf r}_1-{\bf r}_2, \psi  ) =\frac{4}{\pi b^2 \sin  \psi }
\exp\left[- \frac{2{\bf R}_\parallel^2 }{ b^2 \cos^2 (\psi/2)}
- \frac{2{\bf R}_\perp^2 }{ b^2 \sin^2 (\psi /2)} \right]=W_0({\bf R}, \psi  )\,, \label{def_RpRs}
\end{equation}
where $\psi=\phi_1-\phi_2$ is the difference of the orientation angles,  and  ${\bf R}_\parallel = ({\bf r}_1-{\bf r}_2)  \cdot \bar{ \bn}$ and
 ${\bf R}_\perp = -({\bf r}_1-{\bf r}_2)  \times \bar{ \bn}$ are two vectors parallel and perpendicular to the bisector direction $\bar {\bn}$.
The cutoff length $b$ introduced above can be estimated, for example, by comparison of the characteristic kernel width  $\int {\bf R}^2 W_0({\bf R}) d {\bf R} $
for the kernels given by Eqs.~(\ref{rhomb}) and (\ref{kernelb2}) for some typical angle, say $\psi=\pi/2$. Equating both integrals, one finds that $b^2/L^2 = 2/3$.
\footnote{In our previous works \cite{Tsimring_Aranson, Tsimring_Aranson2}
we used a somewhat simpler expression for the kernel, $W_0 ~\sim \exp [ -|\br_1-\br_2|^2/b^2]$.
As we have verified,
this simplified approximation did not change
the results on a qualitative level, affecting only numerical prefactors of some nonlinear terms.}

Finally, the complete kernel can be represented in the form
\begin{equation}
\hspace{-1.1cm}W(\br_1-\br_2,\bn_1,\bn_2)= g W_0 (\br_1-\br_2, \psi)
\left(1+ \frac{\beta}{L} (\br_1-\br_2)\cdot({\bn}_1-{\bn}_2)\right)\,.\,\, \label{kernel1}
\end{equation}
Here $g$ is the interaction rate proportional to the motor density (which can be scaled away) and
the last term $\propto \beta$ describes the anisotropic contribution to the kernel, which is associated
to the increase of motor density towards the polar end of the filament due to dwelling of the motors.
Accordingly, the constant $\beta$ can be related to the dwell time~\cite{Tsimring_Aranson2}.

Near the threshold of the orientation instability mentioned above,
%$\rho \to \rho_c$,
$\rho \gtrsim \rho_c$,
the master equation can be systematically reduced to equations for the coarse-grained local density of filaments
$\rho $ and the coarse-grained local orientation $\tau$
\begin{equation}
\rho = \int_{-\pi} ^\pi P(\br, \phi, t) d \phi\,\,,\quad \btau = \langle \bn \rangle= \frac{1}{2 \pi} \int \bn\,P(\br, \phi, t) d \phi\,,
\end{equation}
by means of  a bifurcation analysis.

%%%%%%%%%%%%%%%%%%%%%%%%%%%%%%%%%%%%%%%%%%%%%%%%%%%%%%%%
\section{\label{crossl}Effects of crosslinks in the model}
%%%%%%%%%%%%%%%%%%%%%%%%%%%%%%%%%%%%%%%%%%%%%%%%%%%%%%%%
\begin{figure}[t]
	\centering
	\includegraphics[width=.6\textwidth]{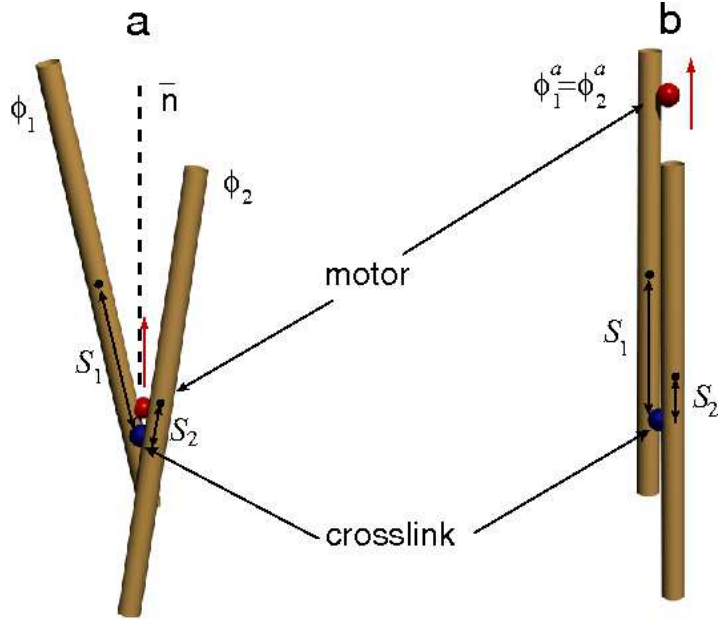}
	\caption{Sketch of the interaction between two filaments, a crosslink and a molecular motor.
	 After the interaction, the motor (shown as a red sphere
	 moving in the direction of the red arrow) aligns the filaments along the bisector $\bar n$, but the midpoint positions do not coincide
	 due to the crosslink (blue sphere).
	\label{fig1}}
\end{figure}

The effect of crosslinks on the motor-induced interaction of filaments is twofold.
First, the simultaneous action of a static crosslink, serving as a hinge,
and a motor moving along both filaments results in a fast and complete
alignment of the filaments, as shown in Fig.~\ref{fig1}.
This justifies the assumption of fully inelastic collisions
for the rods' interaction. Note that without crosslinks the overall change in the relative orientation of the filaments is much smaller:
the angle between filaments decreases only by 25-30 \% in average, see the discussion in Ref. \cite{Tsimring_Aranson2}.
Complete alignment also can occur for the case of simultaneous action of two motors moving in opposite direction, as in experiments  on kinesin-NCD mixtures
reported in Ref.~\cite{Leibler:2001.1}, and
even for two motors of the same type moving in the same direction but with a different speed due to variability of the properties and the
stochastic character of the motion. Second, the crosslinks inhibit relative sliding of rods in the course of alignment,
restricting the motion to rotation only. Thus, in contrast to the situation considered in Refs.~\cite{Tsimring_Aranson,Tsimring_Aranson2}
and described by Eq.~(\ref{wo_crossl}), in the presence of a crosslink the midpoints of the rods {\it will not coincide} after the interaction.
In fact, the distances $S_{1,2}$ from the midpoints to the crosslink point do not change, as it is shown in Fig.~\ref{fig1}.

To describe the interaction rules in the presence of a crosslink,
we express the radius-vector of an arbitrary point on a filament ${\bf
R}_i$ via
the position of its midpoint $\br_i$, the filament orientation
${\bn}_i$, and the distance from the center of mass $S$,
${\bf R}_i  = \bn_i S_i  + \br_i$.
The intersection point of two rods is given by the
condition ${\bf R}^* = {\bf R}_1={\bf R}_2$, which fixes the values of $S_{1,2}$ to
\begin{eqnarray}
S_{1,2}= \frac{(\br_2-\br_1 ) \times \bn _{2,1} }{ \bn _1 \times \bn_2}\,.
\end{eqnarray}
Due to the crosslink, the values of $S_{1,2}$ do not change during the interaction.
Since the filaments become  oriented along the bisector direction  $\bar {\bn}$,
the distance of the two filament midpoints from the total center of mass will be
$\Delta S= (S_1-S_2)$. Therefore, instead of Eqs.~(\ref{wo_crossl_phi}),(\ref{wo_crossl})
we obtain the interaction rules
\begin{eqnarray}
\label{rule_phi}\phi^a_1& =& \phi_2^a  =  \bar \phi=\frac{\phi_1+\phi_2}{2}\,,\\
\label{rule_r}{\br }_{1,2}^a & =&  \frac{ \br_1 + \br_2}{2} \pm \eta\frac { \bar{ \bn} \Delta S     }{2 } = \frac{ \br_1 + \br_2}{2} \pm \eta\frac { \bar{ \bn} ( (\br_1-\br_2)\cdot \bar {\bn})    }{2 \cos \psi }\,.
\end{eqnarray}
Here we have introduced the parameter $\eta$ interpolating between two cases:  the case with crosslinks present corresponds to
$\eta=1$; for $\eta=0$ the previous model, Eqs.~(\ref{wo_crossl_phi}),(\ref{wo_crossl}), is recovered.
Thus the value of $\eta$ can be roughly interpreted as the effective strength of crosslinks or an effective
fraction of crosslinks with respect to motors.

The interaction rules, Eqs.~(\ref{rule_phi}) and (\ref{rule_r}), can be used to evaluate the collision integral, Eq.(\ref{coll_int}).
Omitting lengthy calculations (see the Appendix for details)
after expanding the master equation (\ref{master}) near the
threshold of the orientation instability, we arrive at the following set of nonlinear equations
for the coarse-grained density $\rho$ and orientation $\tau$:
\begin{eqnarray}
\hspace{-2cm} \partial_t \rho &=& D_\rho \nabla^2 \rho-\zeta \nabla^4 \rho
-\frac{\phi_0  B^2 }{64 \pi } \left(1-\eta^2\right)\nabla^2\rho^2\nonumber\\
\hspace{-2cm} & &-\frac{\pi \phi_0 B^2}{16}\left[(S[\phi_0]+ (1-2S[\phi_0])\eta^2)\nabla^2 {\bm \tau}^2+2\left(S[\phi_0]-\eta^2\right) \partial_i\partial_j(\tau_i \tau_j)\right]\,, \label{rho1} \\
\hspace{-2cm} \partial_t \btau&=& \epsilon ( \rho -\rho_c) \btau - A | \btau|^2 \btau + D_{\tau_1}  \nabla^2 \btau + D_{\tau_2}  \nabla \nabla \cdot \btau  +\frac{B^2\rho_0}{4\pi}\nabla^2 \btau \nonumber \\
\hspace{-2cm} & &-H\left[\frac{1}{16\pi}\nabla\rho^2-\left(\pi-\frac{8}{3}\right){\bm \tau}(\nabla\cdot{\bm \tau})-\frac{8}{3}({\bm \tau}\cdot\nabla){\bm \tau}\right]\,,
\label{tau1}
\end{eqnarray}
with $S[x]=\sin (x)/x$ and $B= b/L$.
The constants $A$, $\epsilon$ and the critical density $\rho_c$ are functions of the maximum interaction angle $\phi_0$ and
the inelasticity coefficient $\gamma$:
\begin{eqnarray}
\label{coeff1} \hspace{-2cm} A&=& 2 \phi_0 \frac{\left(S[\phi_0(2\gamma-1)] -S[\phi_0] \right)  \left(S[\phi_0(\gamma+1)]+
S[\phi_0(\gamma-2)]-S(2\phi_0)-S(\phi_0)\right) }{ 2 /  \phi_0 -
(S[2\phi_0(\gamma-1)]+S[2\phi_0\gamma]-S[2\phi_0]-1)\rho/2\pi} \nonumber \\
\hspace{-2cm} \epsilon&=&              \frac{\phi_0} {\pi}  \left[S[\phi_0(\gamma-1)]+
S[\phi_0\gamma]-S(\phi_0)-1\right]\,\,,\quad \rho_c=\frac{1}{\epsilon}\,\,.
\end{eqnarray}
In the following we consider the case $\phi_0=\pi$ as motivated below.
Then the density equation (\ref{rho1}) becomes  somewhat simpler since $S[\phi_0]=0$.
We have introduced rescaled diffusion coefficients, namely
$D_\rho =(D_\parallel+D_\perp)/(2 D_r L^2)=1/32$,
$D_{\tau_1}=(D_\parallel+3D_\perp)/(4 D_r L^2)=5/192$ and
$D_{\tau_ 2}=(D_\parallel - D_\perp)/(2 D_r L^2)=1/96$.
In order to scale out the motor density $g$ we rescaled density and orientation vector, $ g\rho \to \rho$, g$\btau \to  \btau $.
Also length is normalized  by $\br \to \br /L$ and time by $ t \to t/ D_r L^2$.
The anisotropic contribution proportional to $H=\beta b^2/L^2=\beta B^2$ is due to the
polar distribution of the motors along the interacting filaments, while the anisotropic contribution in the $\rho$-equation is due to
the crosslinks.
The isotropic higher order diffusion term $\zeta \nabla^4 \rho$ was included for regularization purposes of the equation
at very short wavelengths.
Assuming additionally $\gamma=1/2$ (i.e. totally inelastic collisions, as justified above),
one obtains from Eqs.~(\ref{coeff1}):
$\epsilon = 4/\pi-1 \approx 0.273$, $A\approx 2.18 $ and the critical density $\rho_c \approx 3.663$.

\begin{figure}[t]
	\centering
	\includegraphics[width=.6\textwidth,angle=-90]{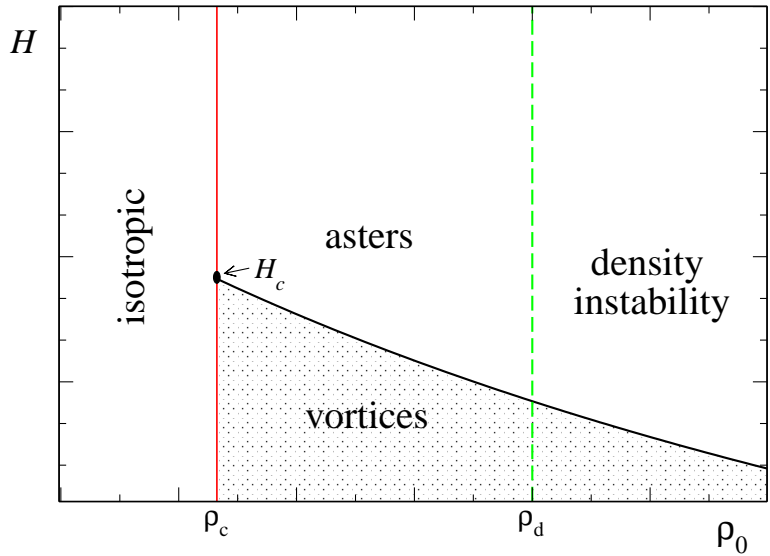}
	\caption{Sketch of the phase diagram of the rescaled density $\rho_0$ (being the product of motor and filament density)
	against anisotropy parameter $H$ in the absence of crosslinks.
	Above $\rho_c$, the polar state is formed.
	Beyond $\rho_{d}$, given by Eq.~(\ref{rhoc1}), an isotropic density instability occurs. Depending on parameters,
	the density instability may happen also prior to the orientation instability, i.e. $\rho_d< \rho_c$.
	In between $\rho_c$ and $\rho_{d}$, asters are stable above the critical line (solid black line)
	while vortices are linearly stable below this line. The critical line terminates at the point $H=H_c$.
	In the presence of crosslinks,  for $\phi_0=\pi$
	bundles occur throughout the polar phases, i.e. beyond the red line.
	However, they are in nonlinear coexistence with the asters/vortices.
	\label{fig2}}
\end{figure}

A sketch of the phase diagram for Eqs. (\ref{rho1}), (\ref{tau1}) in the plane of the motor-induced anisotropy parameter $H$ and the
mean density $\rho_0$ is shown in Fig.~\ref{fig2}.
A uniform isotropic state, $\rho=\rho_0$ and $\btau=0$, loses its stability
if $\rho_0>\rho_c$, independent of the value of $H$. In the spatially uniform case,
 orientation modulations grow
into a polar state with non-zero $|\btau|=[\epsilon(\rho_0-\rho_c)/A]^{1/2}$ and arbitrary orientation of $\btau$.
Recall that the density $\rho_0$ is scaled by the ``collision rate'' $g$, and thus is proportional to both
the density of tubules and the density of motors.
This implies that either increasing the number of motors or the number of filaments can induce the polar phase.
However, in extended systems, the growth of spatially inhomogeneous modes leads to the formation
of a complex  state characterized by disordered arrays of vortices or asters depending on
the value of the anisotropy parameter $H$ \cite{Tsimring_Aranson,Tsimring_Aranson2}.
Vortices are stable only for small values of the
anisotropy parameter $H$; the stability limit of vortices, indicated by the black solid line,
terminates at a critical point at $H=H_c$. The vortex-aster-competition is governed
predominantly by the $\btau$-equation, Eq.~(\ref{tau1}), and
thus prevails whether crosslinks are present or not.
In the case without crosslinks, $\eta=0$,
for densities $\rho>\rho_{d}$, the homogeneous oriented state
loses its stability with respect to density fluctuations as implied by the green dashed line in Fig.~\ref{fig2}.
If crosslinks are present however, $\eta=1$, the density instability is (to leading order) independent of
the value of filament density and thus bundles can be found  throughout the polar phase,
i.e. for all $\rho>\rho_c$, where they are in complicated nonlinear
competition with the aster and vortex defects.

%%%%%%%%%%%%%%%%%%%%%%%%%%%%%%%%%%%%%%%%%%%%%%%%%%%%%%%%
\section{\label{eq_crossl} Instability of the homogeneous polar state}
%%%%%%%%%%%%%%%%%%%%%%%%%%%%%%%%%%%%%%%%%%%%%%%%%%%%%%%%

For $\eta=0$, Eqs.~(\ref{rho1}),(\ref{tau1}) reduce to the model without crosslinks studied  in Refs.~\cite{Tsimring_Aranson,Tsimring_Aranson2}.
 As it was shown in \cite{Tsimring_Aranson,Tsimring_Aranson2},
this equation exhibits an isotropic density instability if $\rho> \rho_{d}$ as calculated below.
In the presence of crosslinks ($\eta=1$), the term in Eq.~(\ref{rho1}) proportional to $\nabla^2\rho^2$ which is responsible
for the density instability
vanishes, and instead a new anisotropic term $\partial_i\partial_j(\tau_i\tau_j)$ appears. This term couples the density and
orientation perturbations already in the linear order. As we will show in the following, this new crosslink-induced
anisotropic coupling
modifies the density instability so it becomes transverse to the direction of polar orientation
(in both the linear and nonlinear regime).

Let us investigate the linear stability of the homogeneous polar solution
of Eqs.~(\ref{rho1}) and (\ref{tau1}), describing a state with density $\rho_0$
and polar orientation $\btau_0$ given by
$\epsilon ( \rho -\rho_c) = A | \btau_0|^2$. Without loss of generality we set
$\btau_0$  along $x$-direction, $\btau_0 = (|\btau_0|, 0)$.
Linearizing the model equations around this state by making
the ansatz $\{\rho,\tau_x,\tau_y\}=\{\rho_0,\tau_0,0\}+\{\delta\rho,\delta\tau_x,\delta\tau_y\} \exp[\sigma ({\bk})
t+  i k_x x + i k_y y]$, one can deduce
the linear growth rates $\sigma$ as a function of the modulation wavenumbers $k_x, k_y$.
 For simplicity we set  $H=0$ here. Finite but small values of $H$
introduce a small drift but only slightly affect the growth rates.

First consider the case without crosslinks ($\eta=0$). Then Eqs.~(\ref{rho1}),(\ref{tau1})
reduce to the model of Refs.~\cite{Tsimring_Aranson,Tsimring_Aranson2}. There are three linear modes in the system.
The two largest growth rates for long-wave perturbations are associated to a transverse orientational mode
and to a mixed density-orientation mode. The third mode, related to the modulus of the orientation, is always damped.
To leading order in $k_x$, $k_y$ the transverse orientational mode reads
\begin{eqnarray}
\sigma_\tau=-\left(D_{\tau_1}+\frac{B^2\rho_0}{4\pi}\right) k_x^2
-\left(D_{\tau_1}+D_{\tau_2}+\frac{B^2\rho_0}{4\pi}\right)k_y^2\,,
\end{eqnarray}
and is thus always damped.
For the mixed density mode one obtains
\begin{eqnarray}
\sigma_\rho=-\left(D_\rho-\frac{B^2\rho_0}{32}\right)(k_x^2+k_y^2)\,.
\end{eqnarray}
Thus a density instability occurs at
\begin{eqnarray}
\rho_0>\rho_d=\frac{32 D_\rho}{B^2}\,,\label{rhoc1}
\end{eqnarray}
as already described in Refs.~\cite{Tsimring_Aranson,Tsimring_Aranson2},
which to leading order is isotropic. Note that depending on the model parameters the density instability for $\eta=0$ may
also occur prior to the orientation instability, i.e. $\rho_d$ can be smaller than $\rho_c$.

A similar analysis can be done in the presence of crosslinks, $\eta=1$.
While the orientational mode remains unchanged,
for the mixed density mode one now obtains
\begin{eqnarray}
\sigma_\rho=-\left(D_\rho+\frac{B^2\pi^2\epsilon}{16 A}\right)k_x^2-\left(D_{\rho}-\frac{B^2\pi^2\epsilon}{16 A}\right) k_y^2\,.
\end{eqnarray}
For perturbations in $x$-direction, i.e. parallel to the polar orientation, %($k_y=0$),
the density mode is damped. However,
using the estimates from above, $\epsilon \approx 0.273, A=2.18, D_\rho =1/32$, and $B^2 \approx 2/3$, one obtains that the coefficient in front of
$k_y$ is negative:
$D_{\rho}-\frac{B^2\pi^2\epsilon}{16 A}<0$, i.e. transverse perturbations (i.e. with small $k_x$ and finite $k_y$)  are {\it unstable}.

Although this linear analysis reveals the possibility of a transverse instability in the presence of crosslinks,
it is not clear if the density modulations perpendicular to the filament orientation really lead
to bundle-like structures in the nonlinear regime. To investigate the
long-term development of this instability, we performed
numerical simulations of the full set of equations (\ref{rho1}),(\ref{tau1}), as described below.

\section{Numerical studies}

\begin{figure}[t]
	\begin{center}
	\includegraphics[width=.85\textwidth]{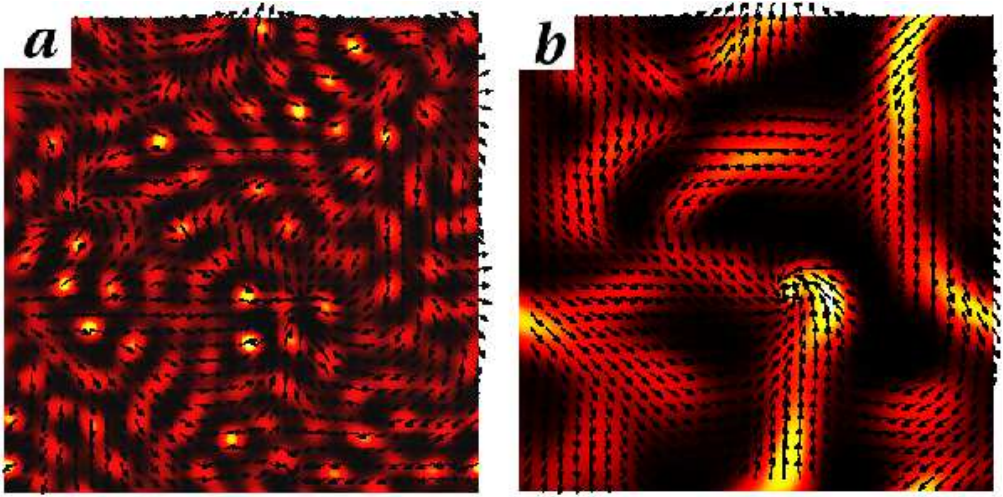}
	\caption{Composite image of the density (color code: black low density, bright yellow high density) and the filament orientation field (arrows).
	a) The model of Ref.~\cite{Tsimring_Aranson2} ($\eta=0$) for $\rho_0=5$ in the region of the isotropic density instability.
	Here the filament orientation is uncorrelated with the density gradient.
	b) The model with crosslinks ($\eta=1$) for a density of $\rho_0=6$ displays pronounced bundles,
	and the local filament orientation is predominantly along the bundles.
	Other parameters values: $H=0.005, B^2=0.6, \zeta=0.04$.
	\label{fig3}}
\end{center}
\end{figure}

In order to study the system beyond the linear regime, we performed numerical investigations of
Eqs.~(\ref{rho1}),(\ref{tau1}).
The studies were conducted in a $80L  \times 80 L$  periodic domain,
for different values of the parameter $\eta$ characterizing the concentration of crosslinks.
Small amplitude noise was used as an initial condition for the $\btau$ field, and $\rho =\rho_0$ $+$ noise for the density field. 
Representative results for $\eta=0, 1$ are presented in Fig.~\ref{fig3}.
In both situations, the simulations were performed in the regime where the homogeneous oriented state
is unstable with respect to density fluctuations.
However, depending on the value of the parameter $\eta$, the
manifestation of the instability is different. For $\eta=0$ (without crosslinks), the numerical solution shows
that the filament orientation and density gradients are mostly uncorrelated, cf. Fig.~\ref{fig3}a.

\begin{figure}[t]
	\centering
	\includegraphics[width=.85\textwidth]{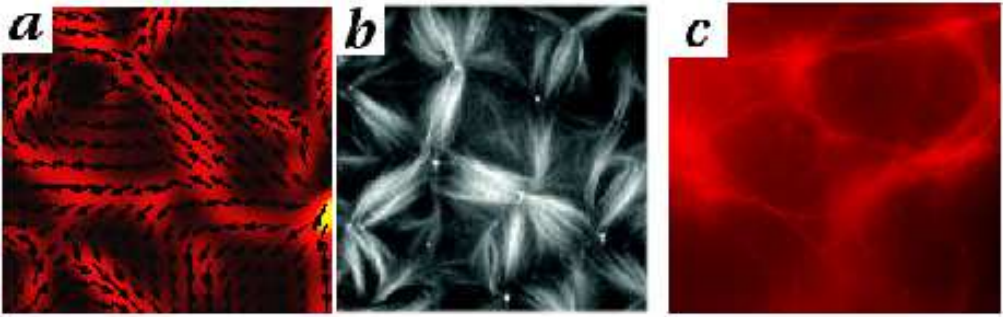}
	\caption{
	a) Composite image of the density (color code: black low density, red high density) and orientation field (arrows)
	for the model with crosslinks, same parameters as in Fig.~\ref{fig3}b but at an early stage of evolution.
	b) Structure observed in microtubule-kinesin-NCD mixtures from {\it Surrey et al.} \cite{Leibler:2001.1}. Here
	the two oppositely directed motors can be effectively mapped to the case of motor and crosslink.
    	c) Experiment on actomyosin by  {\em Smith et al.}, where ATP-depleted oligomeric myosin-motors 
	become crosslinks \cite{Smith:06.1}.
	\label{fig4}}
\end{figure}

In contrast, for $\eta=1$ (with crosslinks),
we observed that the instability indeed resulted in the formation of anisotropic bundles with the filaments' orientation
predominantly along the bundles, as shown in Fig.~\ref{fig3}b. The bundles show a tendency to coarsen with time:
small bundles coalesce into bigger bundles.
The overall pattern is reminiscent of experimental observations of self-organization in both microtubules interacting
with a mixture of motors of two different directions (kinesin and NCD) \cite{Leibler:2001.1}
and experiments on actomyosin where ATP-depleted myosin motors become crosslinks, cf. Fig.~\ref{fig4}.

In order to characterize the degree of alignment quantitatively, we calculated the alignment
coefficient between the density gradient $\nabla \rho$
and the orientation $\btau$:
\begin{equation}
C= 2 \langle \sin^2  (\phi_\rho-\phi_\tau) \rangle -1\,,
\end{equation}
where $\phi_\rho$ and $\phi_\tau$
are the angles between $\nabla \rho$ and the $x$-axis and $\btau$ and the $x$-axis correspondingly.
The alignment coefficient $C=1$ if the vectors $\nabla \rho$ and $\btau$ are everywhere perpendicular,
and $C=-1$ if they are parallel or antiparallel.

We find an alignment coefficient of $C=-0.0045$ for the image
shown in Fig. \ref{fig3}a and corresponding to  $\eta=0$ (no crosslinks), confirming  that the
 fields $\btau$ and $\nabla \rho$ are practically uncorrelated.
For the situation shown in Fig. \ref{fig3}b and corresponding to  $\eta=1$ (crosslinks), we obtained a much larger value of
$C\approx 0.188$, implying that the density gradient and the orientation are predominantly orthogonal.
That means that density modulations are transverse to the orientation within a bundle.

%%%%%%%%%%%%%%%%%%%%%%%%%%%%%%%%%%%%%%%%%%%%%%%%%%%%%%%%
\section{Conclusions}
%%%%%%%%%%%%%%%%%%%%%%%%%%%%%%%%%%%%%%%%%%%%%%%%%%%%%%%%

As we have demonstrated above,
the effect of crosslinks on the organization of polar filaments is twofold.
First, the crosslinks, acting as
hinges, allow zipping and result in the alignment of polar filaments by directional motion of molecular motors.
Second, the ensuing polar state is unstable with respect to
transverse density perturbations yielding bundles of oriented filaments, in contrast to the case without crosslinks in which the density instability is isotropic.

This result has a simple physical interpretation. In the absence of crosslinks the motors tend to bring together the mid-point positions
of microtubules, triggering an isotropic density instability.
This instability is a direct counterpart of the aggregation or  clustering in a gas
of inelastic or sticky particles \cite{Arts}.
With a crosslink holding two filaments together at the intersection point, however, the motion of the filaments along the bisector is suppressed whereas the angular aggregation
proceeds unopposed (furthermore, in fact it becomes much more effective).
Thus crosslinks turn the isotropic instability into a transversal one.

\begin{figure}[t]
	\centering
	\includegraphics[width=0.6\textwidth, angle=-90]{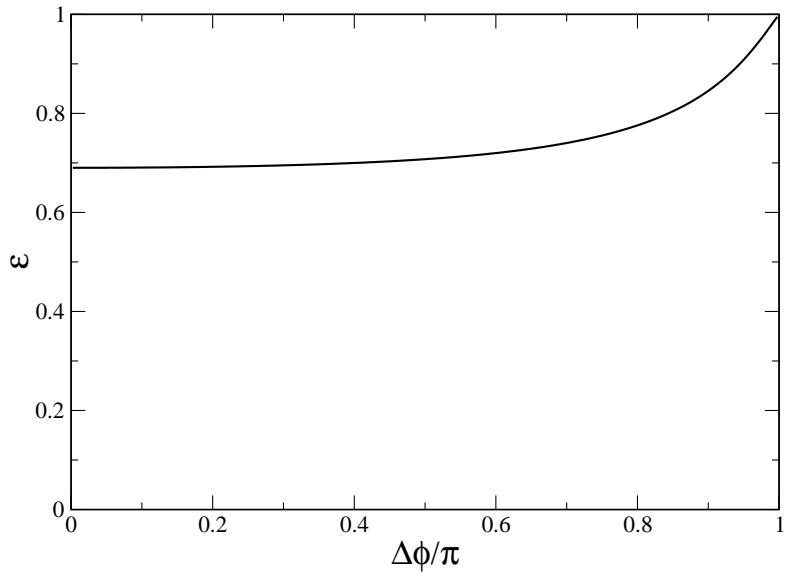}
	\caption{The effective inelasticity factor $\varepsilon$  as a function of the
	initial angle between two rigid filaments $\Delta \phi/\pi$, see for detail  Ref. \protect  \cite{Tsimring_Aranson2}.  }
	\label{fig5}
\end{figure}

There are two experiments related to the model described here.
The experiments on microtubules in the presence of two oppositely directed motor species, as reported in Ref.~\cite{Leibler:2001.1},
appear to produce the same qualitative result as the case of a single motor species mixed with
crosslinks. This is because two motors moving from an initial intersection point
in opposite directions along filaments also lead to their complete alignment.
Additionally, our analysis possibly sheds new light on the interpretation of recent experiments
by Smith {\it et al.} \cite{Smith:06.1}
on actin - myosin mixtures. In this experiment, no patterns were observed in a situation with abundant ATP.
However, long dense bundles of actin filaments were observed when ATP was depleted by the multi-headed myosin
motor constructs, for which
it is known that in the absence of ATP  they rigidly attach to the actin filaments and effectively become static crosslinks.
Also in accordance with this interpretation, reinjection of ATP into the motor-filament solution
resulted in a consequent dissolution of the bundles and
homogenized the system anew.

This experimental result can be interpreted as follows. As mentioned earlier, in the absence of crosslinks,
the interaction of one motor with a filament pair does not result in complete alignment.
In fact, the average decrease of the relative angle is of the order of 25-30\% only,
corresponding to a value of the restitution coefficient of $\gamma\approx 0.85$  or to a value of the effective
inelasticity factor $\varepsilon =2 \gamma-1 \approx 0.7 $. Since the
inelasticity factor approaches 1 at large $\psi$ (see Fig. \ref{fig5}),
it effectively produces a cutoff interaction angle
of the order of $\phi_0 \approx 0.6 \pi$. Filament flexibility
only slightly decreases this value \cite{Karpeev:2007}.
Using the above values for $\gamma$ and $\phi_0$, one finds from
Eq.~(\ref{coeff1}) that the critical density needed for the
orientational instability is about $\rho_{c0}  \approx 15.7$.
However, in the presence of crosslinks, the interaction becomes fully inelastic,
and is described by the restitution coefficient
$\gamma=1/2$. Also, the interaction leads to a complete alignment for
any initial angle, so we can take $\phi_0 =\pi$.
The critical density for these conditions ($\gamma=1/2, \phi_0 = \pi$) is $\rho_{c1} \approx 3.66$, which is more then four
times smaller. Thus in the experiments, even if without crosslinks the motor density was not
high enough to trigger the orientation transition,
due to the crosslinking by ATP-depleted motors the system is likely driven beyond
the threshold of orientation transition. Moreover, the oriented state 
is typically unstable with respect to a transverse instability leading to bundle formation,
implying that bundles are competing with aster-like structures as the experimental pictures suggest.

The inclusion of crosslinks in the model of filament interaction via molecular
motors, Ref.~\cite{Tsimring_Aranson}, was straightforward and yielded nontrivial
results. However, further generalizations of the model are needed. First, instead of the parameter $\eta$
interpolating between the cases with and without crosslinks,
an additional field for the density of crosslinks should be introduced. In case of the actomyosin system,
where ATP-depleted motors are acting like crosslinks, this field might be coupled via some simple
reaction kinetics to the active motor density.
Second, the role of filament flexibility is worth investigating in some detail (cf. \cite{Karpeev:2007}).
Furthermore, it is well known that {\it in vivo}, the cytoskeletal filaments
are often met in a state of constant polymerization and depolymerization by means of ATP and GTP hydrolysis,
another nonequilibrium process that is known to lead to structure formation \cite{Mandelkow:89.1,Hammele:2003.1,Ziebert:2004.1}.
The competition of the two main nonequilibrium processes in the cytoskeleton, active transport of the filaments by molecular motors
and active polymerization of the filaments themselves, might lead to new and surprising behavior.
Finally, an analysis of the homogeneous polar state in a filament-motor model with motor-induced drift,
which we have neglected here, is adressed in Ref.~\cite{Peter:2007.1}.

We thank David Smith and Joseph K\"as
for stimulating discussions and for providing panel c) of Fig.~\ref{fig4}.
This work was supported by the US DOE, grant DE-AC02-06CH11357.

\section{Appendix: Evaluation of the collision integral}
The first term of the collision integral, Eq.~(\ref{coll_int}), can be simplified by integrating out the $\delta$-function
after having expressed $\phi_1$ by
$\phi_1= 2 \phi -\phi_2$ and $\br_1$ by
\begin{equation}
\br_1=\br + \frac{\cos (\psi/2) } {\eta+ \cos (\psi/2) }(\br-\br_2) - \frac{\eta}{\eta+\cos (\psi/2)} \hat A (\br - \br_2)\,,
\end{equation}
where $\psi=\phi_1-\phi_2$ as defined in the main text and where we have introduced the matrix
\begin{equation}
\hat A = \left (
\begin{array}{lr}
\cos (2 \bar \phi) & \sin (2 \bar \phi) \\
 \sin (2 \bar \phi) & - \cos (2 \bar \phi)
\end{array}
\right)\,.
\end{equation}
(Note that after integrating over $\delta(\phi-\bar\phi)$ the angle in the matrix $\hat A$ becomes $\phi$.)
Then one substitutes $w=2 (\phi-\phi_2)$ and $\bxi=\br_1-\br_2$.

In the second term the $\delta$-function leads to $\phi=\phi_1$. After the suitable substitution $w=\phi-\phi_2$
this implies $\bar\phi=\phi-w/2$.
Finally one obtains
the following simple form
\begin{eqnarray}
\hspace{-2cm}\mathcal{I}&=&\int d \bxi \int_{-\phi_0}^{\phi_0} d w
  W(\bxi  ,w ) \left[
 P(\br + \hat A_1\bxi ,\phi+w/2 ) P(\br - \hat A_2 \bxi ,\phi-w/2 )\right. \nonumber \\
\hspace{-2cm}& & \hspace{4.5cm} \Big. -
 P(\br , \phi ) P(\br -\bxi,\phi-w)      \Big]\,,  \label{coll_sub}
\end{eqnarray}
with
\begin{eqnarray}
\hat A_1=\frac{2\cos(w/2)\hat 1-\eta(\hat 1+\hat A)}{4\cos(w/2)}\,\,\,,\,\,\,\hat A_2=\frac{2\cos(w/2)\hat 1+\eta(\hat 1+\hat A)}{4\cos(w/2)}\,.
\end{eqnarray}
In case of $\eta=0$, i.e. in the absence of crosslinks, one regains $\hat A_1=\hat A_2=1/2$ as in the model
of Ref.~\cite{Tsimring_Aranson2}.

To evaluate the spatial integral, one has to transform to the coordinates ${\bf R}=(R_\parallel,R_\perp)$
introduced in the kernel, Eq.~(\ref{def_RpRs}). These are connected to $\bxi$
via a simple rotation,
\begin{equation}
{{\bf R}_\parallel \choose {\bf R}_\perp} = \hat{R}_{\bar\phi}
 { \xi_x \choose \xi_y} \,,\,\quad\hat{R}_{\bar\phi}=\left( \begin{array}{lr}
\cos \bar \phi & \sin \bar \phi \\
-\sin \bar \phi & \cos \bar \phi
\end{array}
\right) \,,
\end{equation}
and the collision integral becomes
\begin{eqnarray}
\hspace{-2cm}\mathcal{I}&=&\int dR_\parallel dR_\perp \int_{-\phi_0}^{\phi_0} d w
  W({\bf R} ,w )\Big[
 P(\br + \hat A_1\hat R_{\phi}\bxi ,\phi+w/2 ) P(\br - \hat A_2 \hat R_{\phi}\bxi ,\phi-w/2 )\Big. \nonumber \\
\hspace{-2cm}& & \hspace{5.6cm} \Big. -
 P(\br , \phi ) P(\br -\hat R_{\phi-w/2}\bxi,\phi-w)\,       \Big]\,. \label{coll_fin}
\end{eqnarray}

%%%%%%%%%%%%%%%%%%%%%%%%%%%%%%%%%%%%%%%%%%%%%%%%%%%%%%%%
\section{\label{References}References}
%%%%%%%%%%%%%%%%%%%%%%%%%%%%%%%%%%%%%%%%%%%%%%%%%%%%%%%%

\end{document}